\documentclass[runningheads]{llncs}
\usepackage[T1]{fontenc}
\usepackage{graphicx}
\usepackage{tabularx}
\usepackage{soul}

\usepackage{amsfonts,amsmath}
\usepackage{algorithmic}
\usepackage{textcomp}
\usepackage{xcolor}

\usepackage[hidelinks]{hyperref}
\usepackage{chngpage}
\usepackage[para]{footmisc}
\usepackage{multirow}
\usepackage{rotating}
\usepackage{lipsum}
\usepackage{booktabs}
\usepackage{caption}
\usepackage{subcaption}
\usepackage{xstring}
\usepackage{color}
\usepackage{cite}

\usepackage{soul}
\usepackage{xcolor}

\definecolor{cb-black}{RGB}{0,0,0}
\definecolor{cb-orange}{RGB}{230,159,0}
\definecolor{cb-skyblue}{RGB}{86,180,233}
\definecolor{cb-bluishgreen}{RGB}{0,158,155}
\definecolor{cb-yellow}{RGB}{240,228,66}
\definecolor{cb-blue}{RGB}{0,114,178}
\definecolor{cb-vermillion}{RGB}{213,94,0}
\definecolor{cb-reddishpurple}{RGB}{204,121,167}

\definecolor{darkgreen}{rgb}{0.0, 0.2, 0.13}

\newcommand{\eg}{\emph{e.g.}}
\newcommand{\ie}{\emph{i.e.}}
\newcommand{\etal}{\emph{et al.}}

\newcommand{\topic}[1]{\hfill\break\noindent\textbf{#1}.}

\newcommand{\weights}{\textrm{\textbf{W}}}

\newcommand{\score}[2]{#1 \footnotesize{$\pm$ #2}}

\newcommand{\mue}{mu2X}

\begin{document}
\title{Exploring Content and Social Connections of Fake News with Explainable Text and Graph Learning}
\titlerunning{Explainable Text-Graph Learning for Fake News Detection}

\author{V\'{i}tor N. Louren\c{c}o\inst{1}\orcidID{0000-0003-0711-1339} \and
Aline Paes\inst{1}\orcidID{0000-0002-9089-7303} \and
Tillman Weyde\inst{2}\orcidID{0000-0001-8028-9905}}
\authorrunning{Louren\c{c}o et al.}
\institute{Universidade Federal Fluminense, Niter\'{o}i, Rio de Janeiro, Brazil
\email{vitorlourenco@id.uff.br}, \email{alineapes@ic.uff.br}\\
City St George's, University of London, United Kingdom\\
\email{T.E.Weyde@city.ac.uk}}


\maketitle              
\begin{abstract}
The global spread of misinformation and concerns about content trustworthiness have driven the development of automated fact-checking systems. Since false information often exploits social media dynamics such as ``likes'' and user networks to amplify its reach, effective solutions must go beyond content analysis to incorporate these factors. Moreover, simply labelling content as false can be ineffective or even reinforce biases such as automation and confirmation bias. This paper proposes an explainable framework that combines content, social media, and graph-based features to enhance fact-checking. It integrates a misinformation classifier with explainability techniques to deliver complete and interpretable insights supporting classification decisions. Experiments demonstrate that multimodal information improves performance over single modalities, with evaluations conducted on datasets in English, Spanish, and Portuguese. Additionally, the framework's explanations were assessed for interpretability, trustworthiness, and robustness with a novel protocol, showing that it effectively generates human-understandable justifications for its predictions. The code and experiments are available at \url{https://github.com/MeLLL-UFF/mu2X/}.

\end{abstract}
\section{Introduction\label{sec:intro}}


The transition from traditional print media to digital platforms has transformed news consumption, with social networks (\eg{}, Reddit, Twitter, Facebook, WhatsApp, TikTok) becoming primary news-sharing platforms. Recent studies demonstrate this shift's magnitude: Pew Research Center's 2021 report\footnote{\url{https://www.pewresearch.org/journalism/2021/09/20/news-consumption-across-social-media-in-2021/}} reveals 48\% of U.S. residents rely on social networks for news, while Statista's Latin America Dossier\footnote{\url{https://www.statista.com/topics/8083/news-in-latin-america}} indicates approximately two-thirds of surveyed individuals in Argentina, Brazil, Chile, and Mexico primarily consume news through social media. This democratisation of information-sharing, while enabling universal content creation, has led to widespread misinformation. The impact is particularly evident in recent elections and the COVID-19 pandemic, where the lack of distinction between traditional and user-generated content has contributed to significant political and social challenges~\cite{Zhou2020,alam2022,Guo2022}.

A prevalent method for addressing misinformation is the use of fact-checking organisations (such as the British Full Fact, the Brazilian Aos Fatos, and the Argentine Chequeado). These organisations typically rely on various sources and documents, usually encompassing images, videos, and text covering a diverse set of topics and international protocols\footnote{\url{https://www.ifcncodeofprinciples.poynter.org/}}.  They employ human assessment of the credibility of social media claims and news articles, which is time-consuming and often a bottleneck in providing verified, trustworthy information.

To address scalability challenges associated with manual fact-checking, several studies~\cite{shu2019,Qi2019,shu2020,hu2021,Verma2021,Shang2021,Qian2021,Singh2021,gupta2022,wu2023,Kawintiranon2023,Panchendrarajan2024} have proposed methods for automatic fact-checking of social media content. 
For instance, \cite{Shang2021}~have introduced a multimodal approach for detecting COVID-19-related misinformation on TikTok. Additionally, \cite{Nielsen2022mumin}~developed a resource known as \texttt{MuMiN}, which establishes connections between claims and Twitter posts. 
These automated approaches aim to streamline the fact-checking process and enhance scalability by providing classification approaches and resources (\eg{}, large-scale datasets) that effectively fill gaps in existing misinformation detection mechanisms. 

Whilst aforementioned automatic fact-checking systems improves scalability, they often lack a crucial component: explainability~\cite{lourencco2022modality,warren2025show,Augenstein2024}, which is essential for user trust and legal compliance. Recent studies propose explainable frameworks for fake news detection~\cite{Kai2019,limeng2019,Kou2022} and multimodal misinformation detection~\cite{Shang2022,Yao2023}. However, none of these approaches integrates multiple languages, content types, and social media features, nor do they provide a robust and comprehensive evaluation of their explainability.

This work introduces \mue{}, a framework for addressing fact-checking in social media posts from \textbf{mu}ltimodal, \textbf{mu}ltilingual, and e\textbf{x}plainable perspectives. 
\mue{} tackles explainable misinformation detection from complete and interpretable aspects, as defined by \cite{Gilpin2018}. 
An \emph{interpretable explanation} describes the system's inner workings in a manner that humans understand; it must generate sufficiently straightforward explanations for an individual to grasp, employing terminology that resonates with the user. 
A \emph{complete explanation} describes the system's inner workings precisely; it must generate explanations that facilitate the anticipation of behaviours across a broader range of scenarios. 
Striking a balance between interpretability and completeness is challenging, since the most precise explanations are frequently less straightforward for individuals to grasp~\cite{Gilpin2018}. Conversely, the most easily understandable explanations often lack the ability to offer a more detailed understanding of the system's behaviour and so to offer predictive insights that inform decision-making.

Our work leverages the explanations from two post-hoc model-agnostic explainability frameworks to \textit{(i)} enhance awareness of the relevant features that led to a classification -- contributing towards completeness; and \textit{(ii)} enrich and combine the resulting explanations regarding the multiple modalities considered -- contributing towards interpretability. 
In particular, our framework incorporates multiple modalities presented in social media posts by combining the posts' local network information, textual content, and related metadata to detect misinformation whilst tackling the well-known lack of explainable classifications. 

\begin{figure*}[htbp]
    \centering
    \includegraphics[width=0.75\textwidth]{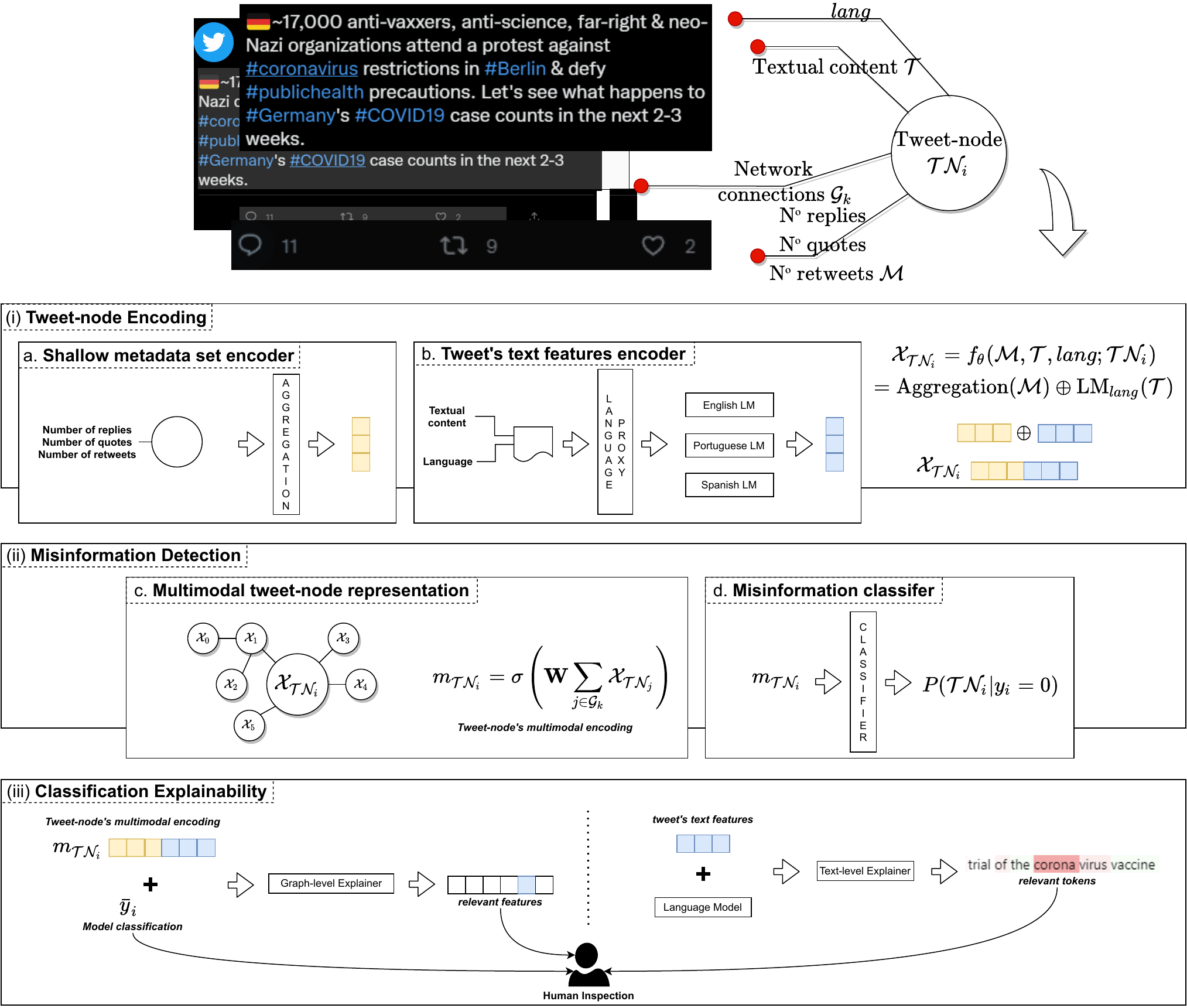}
    \caption{The proposed multi-modal and multilingual explainable framework instantiated to Twitter posts. Yellow and blue elements illustrate graph-based and text-based feature vectors, respectively.}
    \label{fig:framework}
\end{figure*}

As shown in Fig.~\ref{fig:framework}, the proposed framework comprises three main modules: \textit{(i)} Post Encoding Module, where the posts' contents are encoded into multimodal vector representations; \textit{(ii)} Misinformation Detection Module, where the encoded representations are subjected to a graph-based classifier model to classify the posts as factual or misinformation; and \textit{(iii)} Classification Explainability Module, where we employ graph-based and text-based explainable frameworks and combine their outcomes to explain the classification, whilst using the multimodal context to maintain the interpretability of generated explanations.

Our experiments employ the \texttt{MuMiN} dataset, to the best of our knowledge, the only multimodal, multilingual, and multi-topical Twitter-based knowledge base. Our goal is to assess \mue{}'s misinformation detection and explanations capabilities regarding its interpretability, trustworthiness, and robustness. Our findings show that combining different data modalities improves overall misinformation detection, and the explainable setting enables interpretable, reliable, and robust explanations.

\section{Related Work\label{sec:rw}}
In recent years, substantial endeavors have been put into automatically detecting misinformation
\cite{shu2019,Qi2019,limeng2019,Kai2019,kotonya2020,shu2020,Zhou2020,nakamura2020,hu2021,Verma2021,Qian2021,Singh2021,Shang2021,Kou2022,Shang2022,Guo2022,alam2022,gupta2022,Nielsen2022mumin,wu2023,Yao2023,Kawintiranon2023}. 
In the present section, we review methods of misinformation classification concerning the covered modality and explainable capabilities. 
We highlight work that discusses a similar holistic view of misinformation classification in social networks, addressing both the multimodal aspect of information and the explainability of the output classification.

\topic{Misinformation Detection} Most contemporary methods for automated misinformation classification are based on single-modality content, neglecting the multimodal nature of news items. 
For instance, \cite{shu2019,shu2020} address misinformation detection by a method centered around hierarchical propagation networks, \cite{hu2021}~uses graph neural networks supported by external knowledge, \cite{Verma2021} rely on word embeddings over linguistic features, \cite{Kawintiranon2023} use leverage reinforcement learning for fact-checking textual content, and \cite{Qi2019} and \cite{Singh2021} encode visual information to identify fake and manipulated information. 
Some recent work also identifies the benefit of encoding multimodal data to tackle misinformation detection~\cite{alam2022}. \cite{nakamura2020} and \cite{Nielsen2022mumin} propose multimodal (image, text, and relational information) datasets for misinformation detection based on Reddit and Twitter posts, respectively. \cite{gupta2022} introduce a new multimodal and multilingual dataset of misleading articles, training a model that leverages novelty detection and emotion recognition to detect fabricated information. 
\cite{wu2023} propose using multi-reading habits fusion reasoning networks to capture inconsistencies between different types of modalities for news items.

\topic{Explainable Misinformation Detection} The decision-making process of identifying a post or news as misinformation or not goes beyond simply classifying it as such. 
Providing explanations is of foremost importance to prompt human confidence in the classification. 
To that end, \cite{limeng2019} and \cite{Kai2019} propose DEFEND, one of the first explainable methods for detecting fake news articles. 
\cite{kotonya2020} tailor a domain-specific approach for automated fact-checking of claims regarding public health, whilst providing local model-agnostic explainability using LIME~\cite{ribeiro2016}. 
In the same domain, \cite{Shang2021} and \cite{Kou2022} propose explainable methods to detect misinformation related to the COVID-19 pandemic on TikTok's short videos and a knowledge-graph-based crowdsourced dataset. 
Closest to our work, \cite{Shang2022} propose DGExplain, an explainable multimodal framework based on texts and images for COVID-19 misinformation detection. Finally, a multimodal and multi-topical dataset of articles and a model to automate misinformation fact-checking and provide explanations were recently proposed by \cite{Yao2023}. Their explainable approach consists of generating a ruling statement using text evidence, the model prediction, and the input claim and conducting a policy learning process. 

Our work advances existing research by incorporating relational information and relations' metadata, leveraging multiple languages, and being domain-agnostic. We also contrast by considering the interpretability of the explanation, an aspect overlooked in prior research, and focus on social media content classification comprehensibility and explainability.

\section{Method\label{sec:meth}}
This section introduces \mue{}, our multimodal and multilingual explainable framework.
Our primary goal is to offer an end-to-end solution for this task, highlighting key features that explain the classification, thus enabling better human comprehension. This section covers the problem definition. 
Following this, we offer an overview of the components within \mue{}. 
Lastly, we present the implementation of the components forming this framework.

\subsection{Problem Statement\label{sec:problem}}

We define a post-node $\mathcal{P}_i = (\mathcal{T}, \mathcal{G}_k, \mathcal{M}, lang)$, where $\mathcal{T}$ is the textual content; $\mathcal{G}_k$ defines the local $k$-hop network connections of the post, such as replies, quotes, and re-posts; $\mathcal{M}$ is the post's shallow metadata set, which includes, for example, the tweet's owner, the number of likes, the location, etc; and $lang$ is the language in which the tweet was originally written. 
Our goal is for a model $\sigma$ to identify and explain the most representative features within $\mathcal{P}_i$ that lead a post to be classified as \emph{misinformation} (\ie{}, $y_i = \sigma(\mathcal{P}_i) = 0$) or \emph{fact} (\ie{}, $y_i = \sigma(\mathcal{P}_i)= 1$).

\subsection{\mue{}: Multimodal and Multilingual Explainable Framework}

The proposed \mue{} framework comprises three main modules: \textit{(i)} post-node encoding; \textit{(ii)} misinformation detection; and \textit{(iii)} classification explainability. Fig.~\ref{fig:framework} illustrates the instantiated framework, and each module is described in the following.

\topic{Post Encoding Module} This module comprises the encoding of a given post-node $\mathcal{P}_i$ in its vector representation. 
The module is divided into two components to represent two modalities: the post's shallow metadata set $\mathcal{M}_i$; and its textual content $\mathcal{T}_i$. 
The first part obtains the vector representation of the shallow metadata features with the number of replies, quotes, and re-posts (Fig.~\ref{fig:framework}).
a) The second part encompasses a language proxy defined by $lang$ followed by the language-specific text encoders (Fig.~\ref{fig:framework}. 
b) The language proxy acts as a router, assigning the post's textual content $\mathcal{T}$ to a pretrained language-specific model~(LM). 
The pre-trained language-specific model encoder generates a vector representation of the textual content. 
Finally, as described in Equation~\ref{eq:encoding}, both feature vectors are concatenated to form a unique multimodal post-node vector representation $\mathcal{X}_{\mathcal{P}_i}$.

{
\begin{equation}\label{eq:encoding}
\begin{aligned}
    \mathcal{X}_{\mathcal{P}_i} & = f_\theta(\mathcal{M}, \mathcal{T}, lang; \mathcal{P}_i)\\
    & = \text{Aggregation}(\mathcal{M}) \oplus \text{LM}_{lang}(\mathcal{T}).
\end{aligned}
\end{equation}
}

\topic{Misinformation Detection Module} 
This module aims to identify misinformation through a graph-based classifier. It leverages the post's encoded content $\mathcal{X}_{\mathcal{P}_i}$ with its graph connection features $\mathcal{G}_k$ to generate a vector representation called \emph{multimodal post-node representation}. To accomplish this, Equation~\ref{eq:gnn} defines the learning process of the multimodal post-node representation. Given a set of post-node features, $\textbf{X} = \{\mathcal{X}_{\mathcal{P}_1}, \mathcal{X}_{\mathcal{P}_2}, \dots, \mathcal{X}_{\mathcal{P}_N}\}, \mathcal{X}_{\mathcal{P}_j} \in \mathcal{G}_k, \mathcal{X}_{\mathcal{P}_j} \in \mathbb{R}^{|\mathcal{X}_{{P}_i}|}$,

{
\begin{equation}\label{eq:gnn}
    m_{\mathcal{P}_i} = \sigma\left(\weights \sum_{j \in \mathcal{G}_k} \mathcal{X}_{\mathcal{P}_j}\right),
\end{equation}
}
where $\sigma$ is a nonlinear function, and $\weights$ is a linear transformation.

Next, the multimodal post-node representation $m_{\mathcal{P}_i}$ undergoes a classification process, where the probability distribution computed by a softmax function is taken from the representation:

{
\begin{equation}\label{eq:prob}
    P(\mathcal{P}_i | y_i = 0) = \text{softmax}(m_{\mathcal{P}_i}).
\end{equation}
}

\topic{Classification Explainability Module} This module explains the predicted label by employing graph-based and text-based post-hoc explainable methods.

The graph-based method $\mathcal{E}_g$ identifies the most important features within the post-node's multimodal representation $\zeta_{\mathcal{P}_i}$, as displayed in Equation~\ref{eq:eg}:

{
\begin{equation}\label{eq:eg}
    \zeta_{g\mathcal{P}_i} = \text{argmin}\;\;\mathcal{E}_g(\text{softmax}(m_{\mathcal{P}_i}),\;\textbf{X}_{\mathcal{P}_i}),
\end{equation}
}
where $\textbf{X}_{\mathcal{P}_i}$ is the local information network of the target post-node $\mathcal{P}_i$.

The text-based method $\mathcal{E}_t$  generates a scoring vector  $\zeta_{t\mathcal{P}_i}$, where each word in the text is associated with a positive or negative importance score: 
{
\begin{equation}
\label{eq:et}\zeta_{t\mathcal{P}_i} = \mathcal{E}_t(\text{LM}_{lang}, \mathcal{T}),
\end{equation}
}
where $\zeta_{t\mathcal{P}_i} \in [-1,1] 
^{|\mathcal{T}|}$, $\mathcal{T}$ is the textual content of the post-node $\mathcal{P}_i$, and $\text{LM}_{lang}$ is the text encoding method. Finally, both explanatory factors and the classification are provided to the 
user for inspection. 

\subsection{Framework Instantiation\label{sec:inst}}
This section details the mechanisms used to instantiate the previously introduced components illustrated in Fig.~\ref{fig:framework}. Here, we focus on Twitter posts, calling them \emph{tweet-node}s.

\topic{Language Models} To encode our text features, we leverage both pre-trained domain and language specific models. 
For English tweets we use HuggingFace's pre-trained BERTweet\footnote{\url{https://huggingface.co/docs/transformers/model_doc/bertweet}} model~\cite{Nguyen2020BERTweet} as encoder; we used BERTweet.BR\footnote{\url{https://huggingface.co/melll-uff/bertweetbr}} 
for Portuguese tweets and RoBERTuito~\cite{perez2022robertuito} for Spanish tweets.

\topic{Graph-based Classifier} As misinformation classifier, we use the widely used Graph Attention Networks~(GAT)~\cite{velickovic2018graph}, followed by a softmax classifier as described in Equation~\ref{eq:prob}. 
As such, we compute Equation~\ref{eq:gnn} employing a self-attention $\alpha_{kj}$, $\alpha: \mathbb{R}^{|\mathcal{X}_{{P}_i}|}\times\mathbb{R}^{|\mathcal{X}_{{P}_i}|} \rightarrow \mathbb{R}$, mechanism over the tweet-node features:
{
\begin{equation*}
    _{\mathcal{P}_i} = \sigma\left(\weights \sum_{j \in \mathcal{G}_k} \alpha_{kj}\mathcal{X}_{\mathcal{P}_j}\right).
\end{equation*}
}
\topic{Classification Explanation} To identify the key features that contribute to the classification, we used GraphLIME~\cite{Huang2020graphlime}, a nonlinear model-agnostic explanation framework that extends LIME~\cite{ribeiro2016} for Graph Neural Networks. This approach uses Hilbert-Schmidt Independence Criterion~(HSIC) Lasso~\cite{Yamada2014} over $k$-hop neighbour nodes of a given target node. Furthermore, we employ Integrated Gradients~\cite{Sundararajan2017} to assign an importance score to each word of a given tweet's textual content. Integrated Gradients is an axiomatic model that assigns importance scores to individual input features. This approach consists of approximating the integral of the model's output gradients concerning the input.

\section{Experiments\label{sec:exp}}
We conduct experiments to show the capabilities of \mue{}, our proposed multimodal and multilingual explainable framework. 
In our experiments, we measure and assess the framework's capabilities of detecting misinformation, providing robust and trustworthy explanatory factors for these detections, and relating multiple explanatory factors to enhance interpretability. 

We carry out our evaluations on a dataset including multilingual and multi-topical texts and network data that contextualise the texts in the social media environment. 
As far as we know, the MuMiN~\cite{Nielsen2022mumin} dataset is the only dataset that fulfils all requirements of our evaluation. In addition, our focus is on characterising the representation of the textual content and relational features.
The MuMiN dataset is a publicly available dataset of multimodal data from Twitter. MuMiN links tweets covering multiple topics and languages with fact-checked claims, including text, metadata, and visual content from the tweets. 
For our study, we relied on the \emph{MuMiN-small} version of this dataset, which consists of $2,183$ claims and $7,202,506$ tweets. To streamline our analysis, we filtered the dataset to consider tweets written in the three most prevalent languages (English, Portuguese, and Spanish), four types of entities (\emph{Claim}, \emph{Tweet}, \emph{Reply}, and \emph{User}), and six relations (\emph{Posted}, \emph{Mentions}, \emph{Retweeted}, \emph{Quote\_Of}, \emph{Reply\_To}, and \emph{Discusses}).

We developed our framework
using PyTorch~\cite{NEURIPS2019_9015}. 
The full code of the framework, the described instantiation and example notebooks are available for review. The dataset split used was the same as proposed in the MuMiN dataset. As described in Section~\ref{sec:inst}, our misinformation classifier is a GAT to account for the relational information. The classifier was trained using a single Nvidia RTX4070 Mobile GPU, with a learning rate of $0.005$, $16$-dimensional hidden layer, and Adam optimiser~\cite{Kingma2015} for a total of $400$ epochs -- the optimal setting we found in our experiments. 
The textual features were encoded using pre-trained language-specific models. The textual encoding is then submitted to a linear mapping of its original $768$-dimensional space vector to a $812$-dimensional space vector to match the dimensions of the shallow metadata set representation. The experiments in Sections~\ref{sec:exp-inter}-\ref{sec:exp-robu} regard only the multilingual configuration, whilst comparing the various modalities.

\subsection{Explainable Tweet Misinformation Detection}\label{sec:exp-mis-det}
The empirical results of tweet classification are reported in Table~\ref{tab:scores}, in which the model assigns each input information to the \emph{misinformation} or \emph{fact} label. We assess three different inputs: graph-based features, text-based features, and multimodal features, which are the concatenation of both graph-based and text-based features, and measure the obtained F1-score. The reported results are generated by $1000$ bootstrap resampling with $95\%$ confidence interval. 

The results in each language demonstrate that the multimodal feature vector achieves better overall performance in misinformation detection than text encoding alone, confirming the results obtained by Nielsen~\etal{}~\cite{Nielsen2022mumin}. A secondary point is that text-based features outperform graph-based ones in low-resource languages and sparse graphs (\eg{}, Spanish and Portuguese).



\subsection{Interpretability of Explanations}\label{sec:exp-inter}

We further examined the explanatory features through a combination of qualitative and quantitative analyses.


\begin{table}[t]
\footnotesize
\caption{Misinformation classification F1 for each language and modality.}
\label{tab:scores}
\resizebox{\textwidth}{!}{%
\begin{tabular}{c|ccccc}
\toprule
\multicolumn{1}{c|}{\begin{tabular}[c]{@{}c@{}}GAT's input\\\footnotesize{Feature representation}\end{tabular}} & Multilingual & English & Spanish & Portuguese \\ 
\midrule
Graph-based & \score{0.9488}{0.0032} & \score{0.9419}{0.0045} & \score{0.9477}{0.0029} & \score{0.9920}{0.0038} \\
Text-based & \score{0.9487}{0.0032} & \score{0.9409}{0.0044} & \score{0.9486}{0.0028} & \score{0.9923}{0.0042} \\
\textbf{Multimodal} & \textbf{\score{0.9738}{0.0057}} & \textbf{\score{0.9738}{0.0053}} & \textbf{\score{0.9500}{0.0004}} & \textbf{\score{0.9951}{0.0053}} \\
\bottomrule
\end{tabular}
}
\end{table}

\begin{figure*}[htbp]
    \centering
    
    \begin{minipage}{\textwidth}
        \captionof{table}{Report of two inferred cases with their classification, features, and explanatory factors identified by our framework.}
        \label{tab:investigation}
        \resizebox{\linewidth}{!}{%
        \begin{tabular}{@{}llccl@{}}
        \toprule
        \multicolumn{1}{c}{Text} &
          Shallow metadata set &
          Classification &
          \begin{tabular}[c]{@{}c@{}}GraphLime\\ most representative features\end{tabular} &
          \multicolumn{1}{c}{Human interpretation} \\ \midrule
        \begin{tabular}[c]{@{}l@{}}Great news! Carona virus vaccine ready. Able to cure patient within 3 hours\\ after injection. Hats off to US Scientists. Right now Trump announced that\\ Roche Medical Company will launch the vaccine next Sunday, and millions\\ of doses are ready from it !!! VIA: @wajih79273180 https://t.co/BZJCLtwuXq\end{tabular} &
          \begin{tabular}[c]{@{}l@{}}Number of retweets 26\\ Number of replies 42\\ Number of quotes 7\end{tabular} &
          Misinformation &
          \begin{tabular}[c]{@{}c@{}}Number of retweets\\ Number of replies\end{tabular} &
          \begin{tabular}[c]{@{}l@{}}Users who retweet tend\\ to spread misinformation\end{tabular} \\
          & & & & \\
        \begin{tabular}[c]{@{}l@{}}17,000 anti-vaxxers, anti-science, far-right \&amp; neo-Nazi organizations\\ attend a protest against \#coronavirus restrictions in \#Berlin \&amp; defy\\ \#publichealth precautions. Let's see what happens to \#Germany's \#COVID19\\ case counts in the next 2-3 weeks. https://t.co/TD5xIoT5sV\end{tabular} &
          \begin{tabular}[c]{@{}l@{}}Number of retweets 11\\ Number of replies 9\\ Number of quotes 2\end{tabular} &
          Fact &
          Text &
          \begin{tabular}[c]{@{}c@{}}Word importance\\illustrated in Fig.~\ref{fig:textexplainer}.\end{tabular}
          \\ \bottomrule
        \end{tabular}
        }
    \end{minipage}
    
    
    \begin{minipage}{\textwidth}
        \centering
        \includegraphics[width=\textwidth]{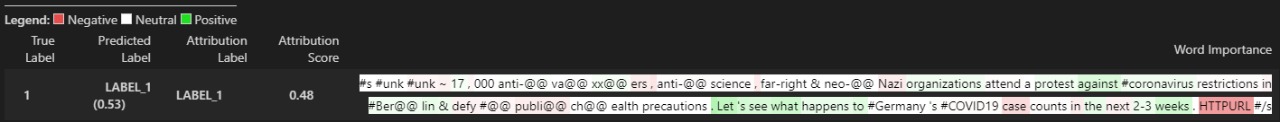}
        \caption{Word importance for text explainability of the second tweet from the table above.}
        \label{fig:textexplainer}
    \end{minipage}
\end{figure*}

Our qualitative examination focused on the most relevant features and their ability to establish a connection between the tweet-node and its associated label. The results of our investigations are presented in Table~\ref{tab:investigation}, where we showcase the tweet-node text and a limited set of metadata, along with the classification produced by our GAT model, the most relevant features identified by GraphLime~\cite{Huang2020graphlime}, and our interpretation of these explanatory factors.

In Table~\ref{tab:investigation}, the  first entry is classified by the GAT model as misinformation. GraphLime highlighted the number of retweets and the number of replies as the most relevant features. 
Upon analysing these retweets and replies, we observed that users who engage in replies or retweets tend to have other tweets classified as misinformation, indicating a trend in the spread of misinformation. 
The second entry in Table~\ref{tab:investigation} was classified as factual, with GraphLime's most relevant feature being the text embedding. 
In addition, we explored textual factors using Captum~\cite{kokhlikyan2020captum} to assist in explaining the label (Fig.~\ref{fig:textexplainer}). 
We noted that phrases like ``protest against \#coronavirus'' and ``Let's see what happens to \#Germany'' contributed to the accurate classification.

\begin{figure}[t]
    \centering    
    \begin{subfigure}{0.45\columnwidth}
        \includegraphics[width=\columnwidth]{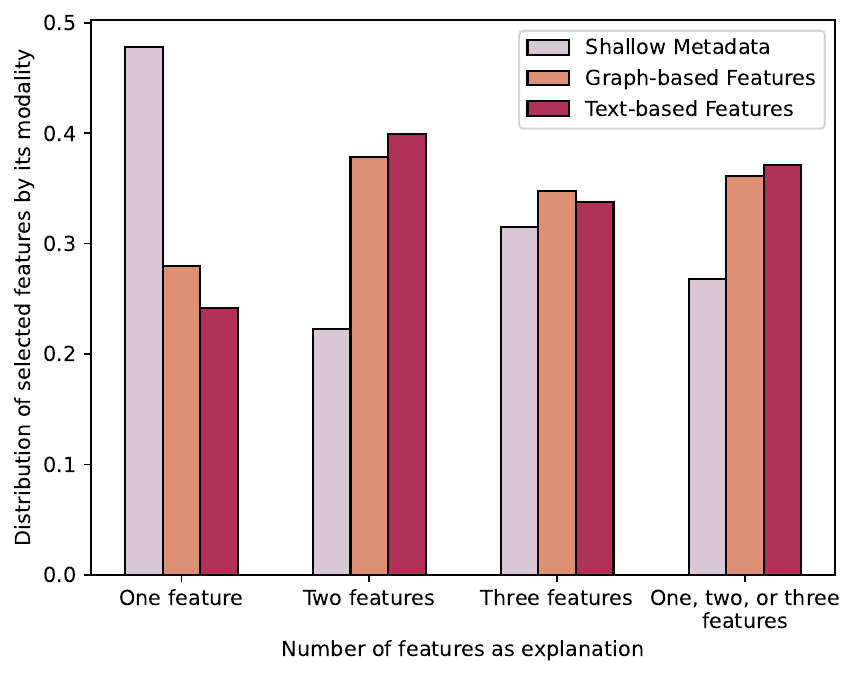}
        \caption{Multilingual}
    \end{subfigure}
    \hfill
    \begin{subfigure}{0.45\columnwidth}
        \includegraphics[width=\columnwidth]{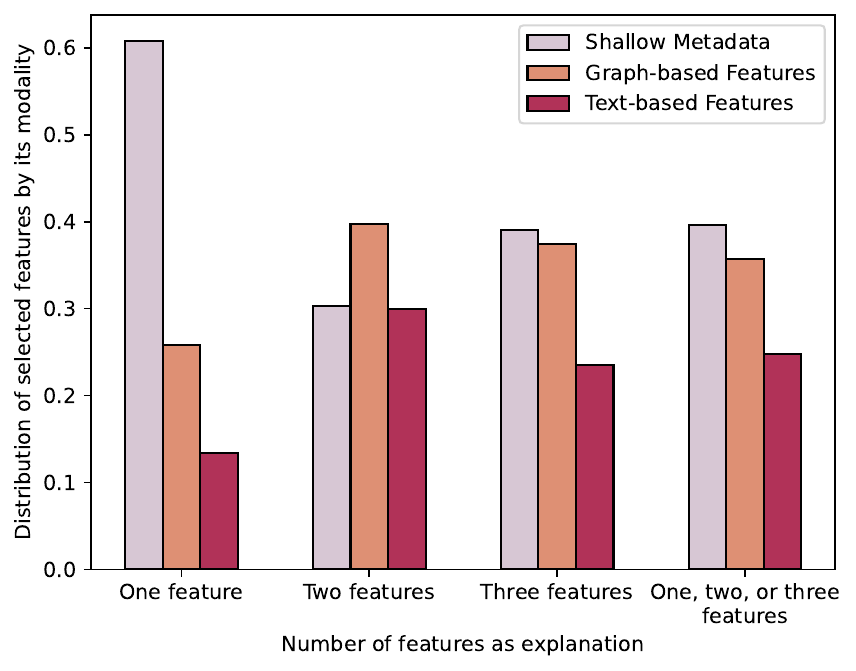}
        \caption{English}
    \end{subfigure}
    \begin{subfigure}{0.45\columnwidth}
        \includegraphics[width=\columnwidth]{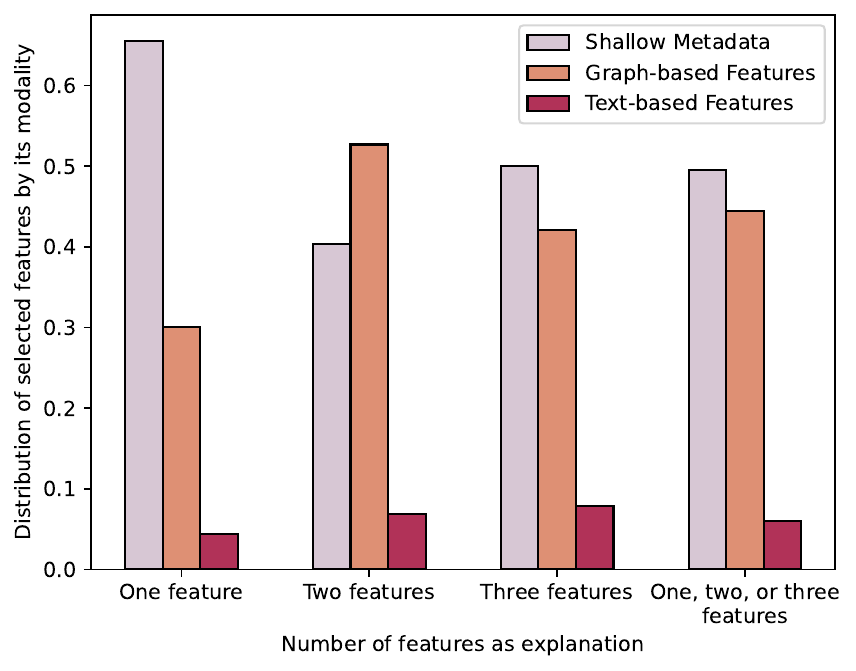}
        \caption{Spanish}
    \end{subfigure}
    \hfill
    \begin{subfigure}{0.45\columnwidth}
        \includegraphics[width=\columnwidth]{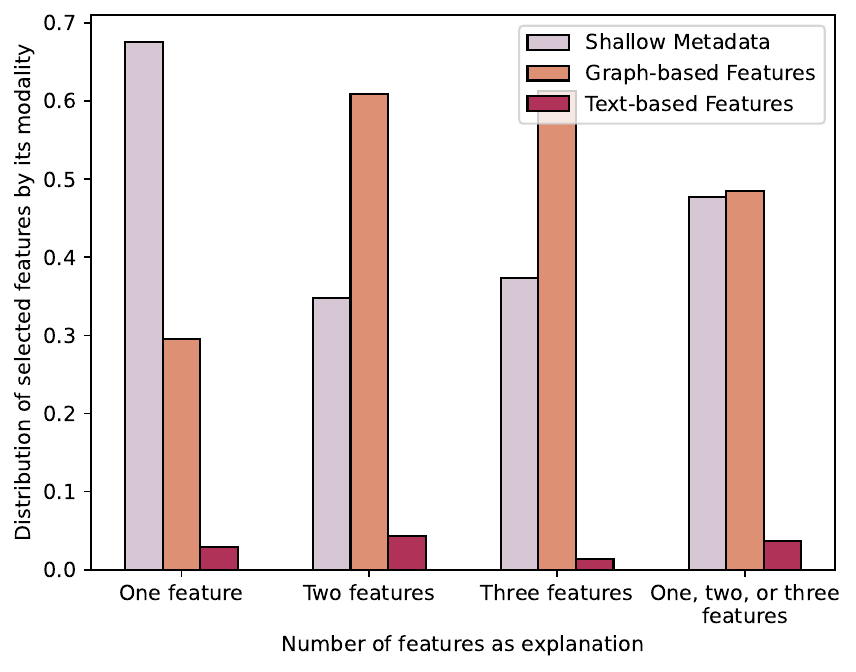}
        \caption{Portuguese}
    \end{subfigure}
    \caption{Distribution of features selected as explanations by their modality. Results are separated by language configuration, and colour indicates each set of features.}
    \label{fig:interstats}
\end{figure}

For our quantitative analysis, we fixed the multimodal setting to study how language and dataset size influence explanatory factors. Fig.~\ref{fig:interstats} shows, for each language, the distribution of features selected as explanations by modality (metadata, graph-based, or text-based) and by the number of features selected (one, two, or three) and the overall (features together). For languages with fewer examples, like Spanish (Fig.~\ref{fig:interstats}.c) and Portuguese (Fig.~\ref{fig:interstats}.d), metadata and graph-based features are used more frequently, followed by text-based features. This is likely due to fewer nodes and less dense relational connections, making these features more discriminative. In contrast, the multilingual setting (Fig.~\ref{fig:interstats}.a) relies more on text-based features, likely due to its denser and more complex tweet-node network. English (Fig.~\ref{fig:interstats}.b) shows a balanced use across modalities, reflecting its intermediate node count and density. Regardless of language, metadata features are the most common explanatory factor when only one feature is selected, aligning with Table~\ref{tab:investigation}'s first entry and reinforcing that reply and retweet counts signal misinformation bubbles.

\subsection{Trustworthiness of Explanations}\label{sec:exp-trust}

In real-world applications, it is fundamental that the classifier's predictions are credible, \ie{}, the classifier should be consistently accurate in associating labels so that the end-user can relate to the prediction as trustworthy. 
Moreover, explanations play an essential role in assessing predictions. Therefore, we develop a protocol to measure the extent to which the explanations have the capacity to assist users in determining the reliability of the prediction.

\begin{figure*}[thbp]
    \centering
    
    \begin{minipage}{\textwidth}
        \centering
        \captionof{table}{F1-Score of \mue{}'s trustworthiness based on different modalities.}
        \label{tab:trustworthiness}
        \resizebox{\textwidth}{!}{%
        \begin{tabular}{cccccc}
        \toprule
        \multirow{2}{*}{Modality} & \multicolumn{5}{c}{F1-Score given top-$k$ most relevant features}                                                                                                                                                                  \\
                                  & Top-$1$                                     & Top-$2$                                     & Top-$3$                                     & Top-$5$                                     & Top-$10$                                    \\ \hline \\
        Graph-based               & 0.7576 \footnotesize{$\pm$ 0.2199}          & 0.8059 \footnotesize{$\pm$ 0.0158}          & 0.8035 \footnotesize{$\pm$ 0.0188}          & \textbf{0.8076 \footnotesize{$\pm$ 0.0137}} & \textit{0.8031 \footnotesize{$\pm$ 0.0187}} \\
        Text-based                & \textit{0.8188 \footnotesize{$\pm$ 0.0164}} & \textbf{0.8096 \footnotesize{$\pm$ 0.0166}} & \textbf{0.8111 \footnotesize{$\pm$ 0.0154}} & 0.8060 \footnotesize{$\pm$ 0.0156}          & 0.8029 \footnotesize{$\pm$ 0.0118}          \\
        Multimodal                & \textbf{0.8207 \footnotesize{$\pm$ 0.0185}} & \textit{0.8061 \footnotesize{$\pm$ 0.0153}} & \textit{0.8080 \footnotesize{$\pm$ 0.0203}} & \textit{0.8062 \footnotesize{$\pm$ 0.0167}} & \textbf{0.8115 \footnotesize{$\pm$ 0.0175}} \\\bottomrule
        \end{tabular}%
        }
    \end{minipage}
    
    
    \begin{minipage}{\textwidth}
        \centering
        \includegraphics[width=0.45\textwidth]{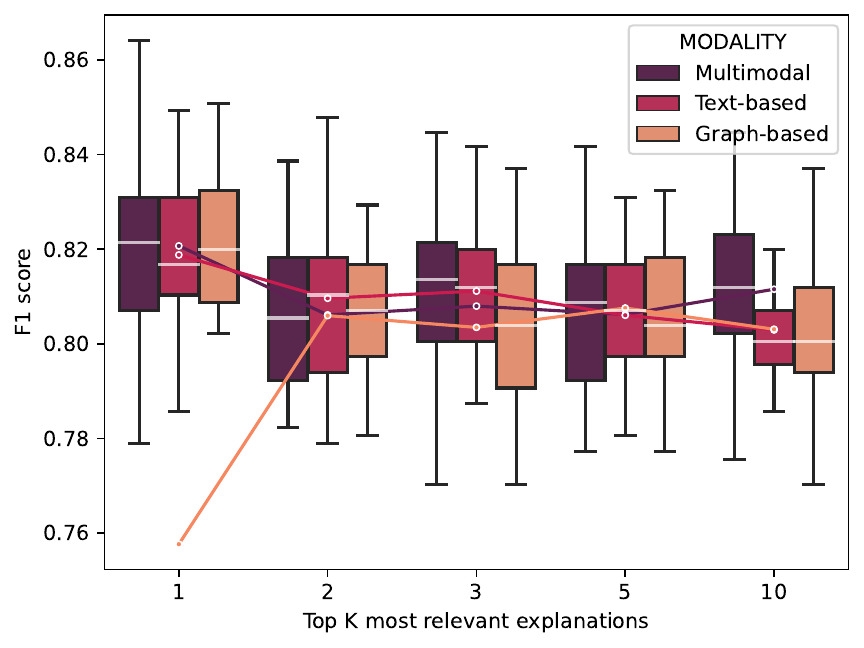}
        \caption{F1-Score of trustworthiness box-plot visualisation. Lines are the average F1-Score in each Top-$K$.}
        \label{fig:trustworthiness}
    \end{minipage}
\end{figure*}

The protocol consists of, first, simulating the trust in individual predictions. 
For that, we randomly select $30\%$ of the features and assume these as ``untrustworthy'', \ie{}, the user would likely recognise these features and would not want them as explanations. 
The second part of the protocol embraces the constructions of an \emph{oracle} ``trustworthiness''. 
We labelled the classifier's prediction as \emph{untrustworthy} if the prediction changes after removing the features selected as untrustworthy from an instance, and \emph{trustworthy} otherwise. 
The oracle assumes that if removing untrustworthy features changes a prediction, those features are decisive for the classifier. In this simulated setup, the oracle serves as the true label for evaluating prediction credibility. The third step simulates user judgment by considering a prediction untrustworthy if it changes after removing all untrustworthy features identified in the explanation using a linear approximation model. Finally, we compare the oracle's label with the simulated users' decisions.

Given the trustworthiness evaluation protocol, we report and compare the F1-score on the trustworthy predictions for the classifier trained with graph-based features, text-based features, and multimodal features over 25 rounds, considering the top-$K$ most relevant explainable features ($K \in \{1, 2, 3, 5, 10\}$). The results are compiled in Fig.~\ref{fig:trustworthiness} and Table~\ref{tab:trustworthiness}. The multimodal classifier achieved the most stable results, having the best F1-score in top-$1$ and top-$10$, whilst being the second-best in the other cases. The classifier that was solely trained with graph-based features displayed the worst results, indicating that it usually mistrusts trustworthy predictions, whilst trusting untrustworthy predictions too often. Despite the used modality, the results show F1-Score above $80\%$ (except the graph-based top-$1$), confirming the reliability of the \mue{} explanations.

\subsection{Robustness of Explanations}\label{sec:exp-robu}

Besides credible predictions, robust explanations are key for the success of misinformation classifiers. We propose a protocol to evaluate the capacity of the explainable method to select relevant features, \ie{}, non-noisy features. Firstly, we artificially added $N = |\mathcal{X}| * p$ amount of randomly generated features, \ie{}, \emph{noisy features}, where $N \in \mathcal{N}$, $|\mathcal{X}|$ is the number of features in a sample $\mathcal{X}$, and $p \in \{0.01, 0.1, 0.25, 0.5, 0.75, 1.0\}$ is the the overall proportion of noisy features to be introduced in each sample. Then, we train the classifier with the added noisy features. Finally, we observe the amount of noisy features selected as explanations.

\begin{figure*}[htbp]
    \centering
    \begin{subfigure}[t]{.33\textwidth}
        \includegraphics[width=\textwidth]{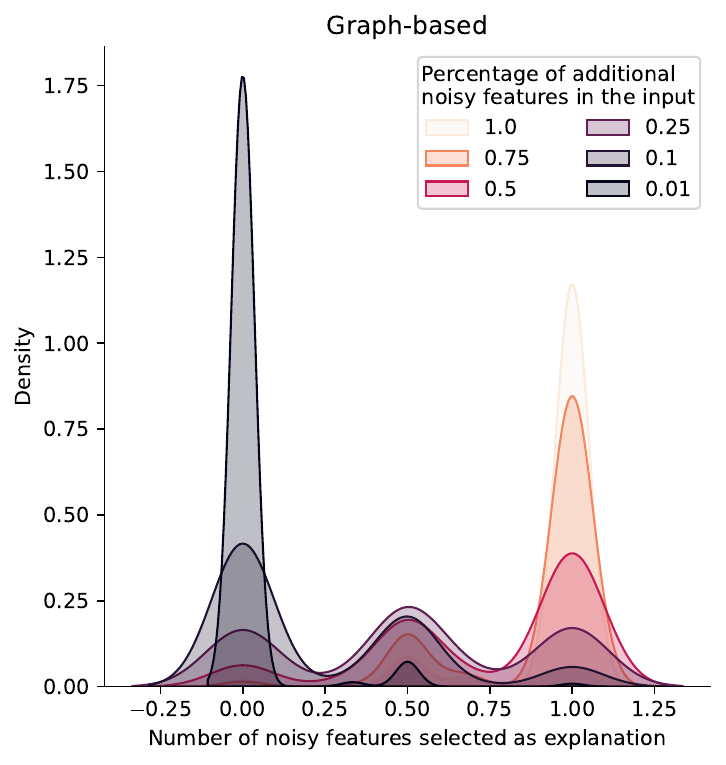}
        \caption{Graph-based features}
    \end{subfigure}%
    \hfill
    \begin{subfigure}[t]{.33\textwidth}
        \includegraphics[width=\textwidth]{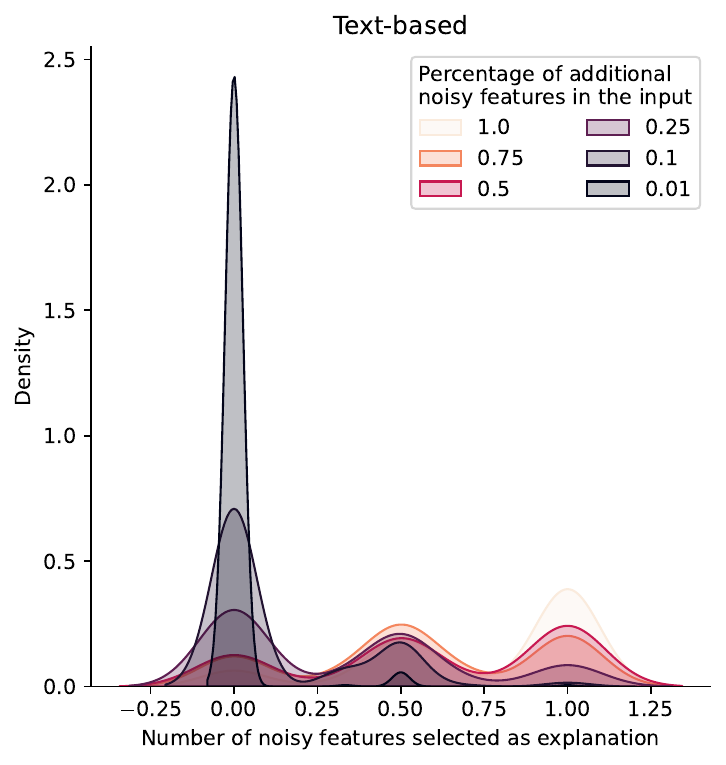}
        \caption{Text-based features}
    \end{subfigure}%
    \hfill
    \begin{subfigure}[t]{.33\textwidth}
        \includegraphics[width=\textwidth]{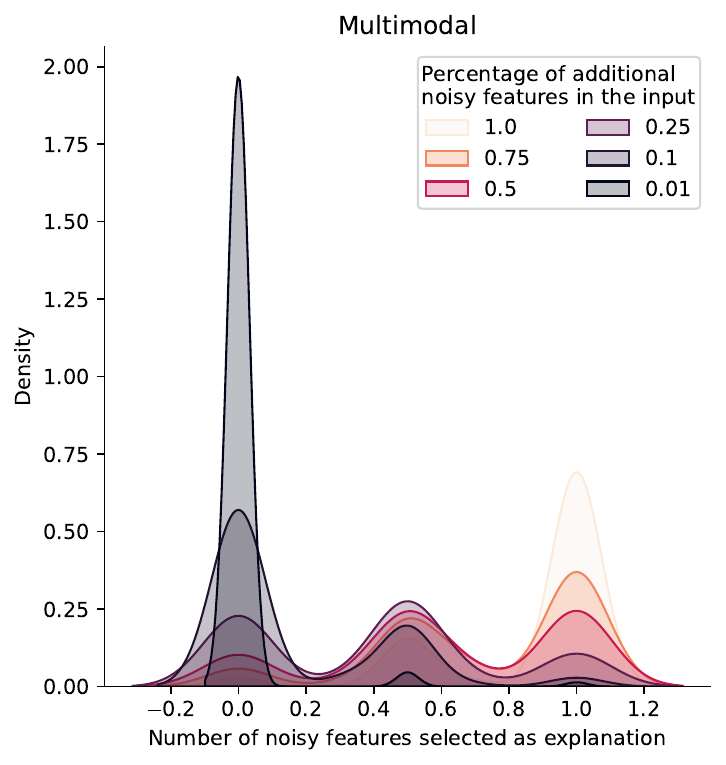}
        \caption{Multimodal features}
    \end{subfigure}
    
    
    \begin{subfigure}[t]{\columnwidth}
        \centering
        \includegraphics[width=0.5\columnwidth]{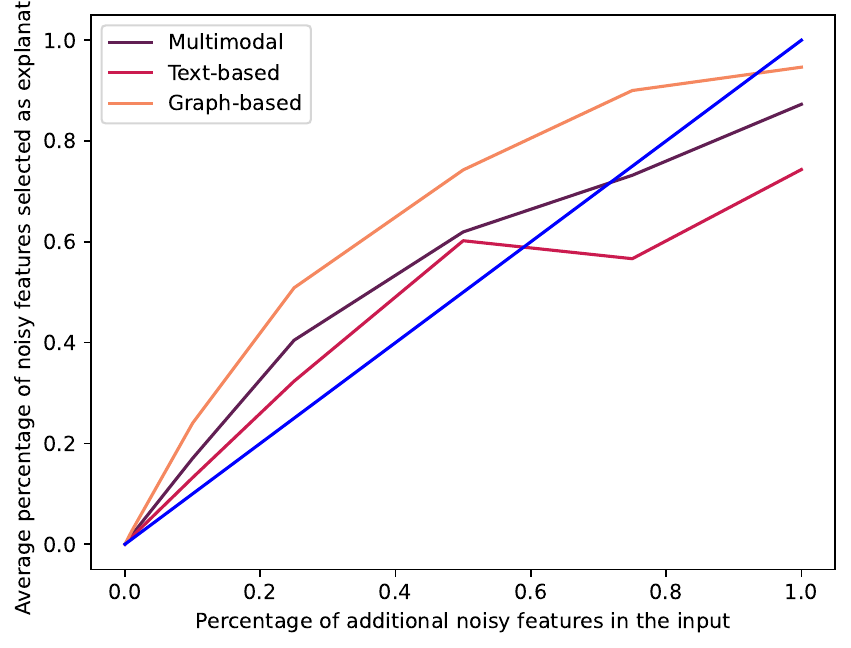}
        \caption{Relation between the percentage of additional noisy features and the average percentage of noisy features selected as an explainable factor.}
        \label{fig:robustness}
    \end{subfigure}
    
    \caption{Distribution of noisy features selected as explanation given the percentage of noisy features added.}
    \label{fig:dist-roubustness}
\end{figure*}

Given the robustness evaluation protocol, we report the average results of 25 rounds of the noisy features selected as explanation patterns for the three classifiers. In Fig.~\ref{fig:dist-roubustness}, we show the kernel density estimation of the frequency of selected noisy features given the classifier and the proportion of noisy features added. The explanations based on the text-based features classifier were more robust, having a high density at zero noisy features selected in the lower percentages and even curves at zero or one noisy feature selected in higher noisy environments. 
The explanations grounded on the multimodal features classifier showed average results, where it keeps the higher density at zero for lower noisy environments and is even out on higher noisy environments.

These patterns can also be observed in Fig.~\ref{fig:robustness}. The figure shows the distribution between the average percentage of noisy features selected as explanations and the percentage of noisy features added. Here, it is observed that the features selected as explanations for the text-based classifier increase linearly with the percentage of noisy features added. Additionally, it can be observed that the features selected as explanations using the multimodal classifier soften the non-linear curve presented by the explanations over the graph-based classifier.



\section{Conclusion\label{sec:conc}}
This paper introduced \mue{}, a multimodal and multilingual explainable framework for misinformation detection in social media. It leverages multiple data modalities and two explainability methods to detect misinformation while clarifying the features that support each classification. Our experiments show that multimodal features enhance detection performance. The proposed explainable methodology adds completeness to the framework by providing trustworthy, robust, and relevant features, along with interpretability, particularly when combined with multimodal data.

There are multiple directions for future work. First, we aim to enhance and further explore \mue{} by incorporating additional modalities and modality-specific explainability methods. We hypothesize that leveraging more modalities will improve both classifier performance and the completeness and interpretability of explanations. Second, we plan to develop a protocol for quantitatively evaluating interpretability based on human assessment, since existing ones do not handle multimodality~\cite{hase2020,colin2022}. Lastly, we envision integrated explanations that combine content snippets with explanatory factors or incorporate prior knowledge to improve clarity and user understanding.

\topic{Limitations} Our work has two main limitations: framework generalization and user interpretability. First, although it supports various multimodal aggregation methods, we only tested vector concatenation. Exploring other strategies could improve performance. Second, the feature-based explainability may be hard for some users to interpret. Still, our approach balances accuracy and interpretability by offering combined explanations across modalities, aligning with the trade-off discussed in~\cite{garouani2024investigating}.

\topic{Declaration on Use of Generative AI}
The author(s) used Claude and Amazon Nova family models for grammar and spelling checks during the preparation of this work. After reviewing and editing the content as needed, the author(s) take full responsibility for the publication's content.

\paragraph*{\textbf{Funding}}
The second author thanks the grants from CNPq~(National Council for Scientific and Technological Development), FAPERJ -- \textit{Funda\c{c}\~{a}o Carlos Chagas Filho de Amparo \`{a} Pesquisa do Estado do Rio de Janeiro}, processes SEI-260003/002930/2024, SEI-260003/000614/2023, and CAPES.

\bibliographystyle{splncs04}
\bibliography{main-short}

\end{document}